\RequirePackage{fix-cm}
\documentclass[smallextended]{svjour3} 
\smartqed 
\usepackage{ams}
\usepackage{calligra}
\usepackage{graphicx}
\usepackage[misc]{ifsym}
\usepackage{float}
\begin{document}

\title{Cryptanalysis of The Quantum Secure Direct Communication and Authentication Protocol With Single Photons
}


\author{Ali Amerimehr \and
Massoud Hadian Dehkordi 
}


\institute{A. Amerimehr , M. H. Dehkordi (\Letter) \at
Department of Pure Mathematics, School of Mathematical Sciences, Iran University of Science and Technology, \\
Narmak, Tehran, Iran \\
\email{ali\_amerimehr@iust.ac.ir} 
\and
M.H. Dehkordi \\
\email{mhadian@iust.ac.ir}
}

\date{Received: date / Accepted: date}

\maketitle

\begin{abstract}
We analyze the security of a quantum secure direct communication protocol equipped with authentication. We first propose a specific attack on the protocol by which, an adversary can break the secret already shared between Alice and Bob, when he (adversary) runs the protocol few times. The attack shows that there is a gap in authentication procedure of the protocol, and by doing so the adversary can obtain the key without remaining any trace of himself. We then give the modification of the protocol and analyze the security of it, and show how the modified protocol can close the gap.
\end{abstract}

\section{Introduction}
\label{intro}
Over the last decades, quantum cryptography plays a significant role in abstract theory of information and communication security. It is divided into some major research topics, such as QKD \footnote{quantum key distribution}, QSS \footnote{quantum secret sharing}, QIA \footnote{quantum identity authentication}, etc. which have developped from their firs publications [1-3] respectively, untill now.

A new topic of quantum cryptography which has been studied at depth, comprehensively recently, is quantum secure direct communication, known as QSDC in the literature. The goal of QSDC is to convey a secret message directly without a key generating session to encrypt the message. Like many other topics of quantum cryptography, there are two approahces to research on QSDC: quantum entanglement \cite{4}, \cite{5}, \cite{r-6}, \cite{r-5}, and single photons \cite{6}, \cite{7}, \cite{8}, \cite{r-7}, \cite{r-12}, \cite{r-9}. 

Recently many papers have been published in QSDC and some related topics such as quantum dialogue (QD) and quantum secure direct dialogue (QSDD).  In 2013, Chang et al. proposed a QSDC protocol equipped with authentication \cite{8}. Chou et al. proposed a bidirectional QSDC protocol for mobile network \cite{chou}. In 2014 Lai et al. proposed a quantum direct secret sharing (QDSD) scheme using fountain codes for eavesdropping check and authentication \cite{r-8}; a pre-shared sequence of degrees and positions is applied to recognize the additional qubits. Hwang et al. introduced a new topic in quantum cryptography called quantum authencryption, combining quantum encryption and quantum authentication into one process for off-line communications \cite{r-7}. Yang put forward a QSDC protocol without quantum memory; a stream is replaced by quantum data block to transmit quantum states \cite{r-12}. Chang et al. proposed a controlled QSDC protocol; they used five-particle cluster state and quantum one time pad \cite{r-6}. Zou and Qiu introduced a semiquantum secure direct communication protocol with classical Alice \cite{r-9}. Gao analyzed the protocol proposed in \cite{chou} and suggested a possible improvement of it \cite{r-2}. In 2015, to combat collective-dephasing noise and collective-rotation noise, Ye put forward two QD protocols \cite{r-1} and a QSDD protocol \cite{r-3}. Xiao and Xu proposed a high- capasity quantum secure communication scheme using either entangled pairs and an auxiliary single photon \cite{r-4}. Hassanpour and Houshmand put forward a three-party controlled QSDC based on GHZ-like states which improves the efficiency of the previous ones \cite{r-5}. Ma et al. presented a direct communication protocol of quantum network over noisy channel by which the bit-flip errors whould be corrected using a parity matrix \cite{r-11}. Chang et al. put forwarda controlled deterministic secure quantum communication protocol with limited bases for preparing and measurement of qubits \cite{r-13}. An experimental implementation of proposed protocol in \cite{5}, is introduced by Hu et al. \cite{r-14}. Mi et al. presented a QSDC scheme using orbital angular momentum of photons to reach higher capacity and security \cite{r-10}. In 2016, Uhlmann found that the anti-(conjugate) linearity plays an importatnt role in the security of quantum cryptographic protocols \cite{r-15}. Li and Yin suggested feasible quantum information processing in terms of living cells which may be applied to experimental demonstration of quantum systems \cite{r-16}.

As mentioned, Chang et al. put forward a QSDC and authentication protocol based on single photons, in which it is assumed that Alice and Bob have two secret strings $\mathrm{ID_{A}}$ and $\mathrm{ID_{B}}$. Analysis of the protocol shows that it is immune against most attcks such as man-in-the-middle attack and quantum teleportation attack \cite{8}. The current article, discovers a new attack on the recent protocol to reaveal the secret information. The attack scenario is divided into two scenes. At the first scene, the attacker who is called ``Oscar" tries to obtain $\mathrm{ID_{B}}$ by sending some separate messages to Bob. At the second scene, he tries to obtain $\mathrm{ID_{A}}$ by braeking a simple XOR-encryption. 

The rest of the article is as follows. Section 2 reviwes Chang et al.'s protocol without any example. For a complete picture of the protocol and also related examples, we refer the readers to see \cite{8} and refrences therein. Section 3 is devoted to present the new attack. In section 4, the accuracy of the new attack is studied, and the probability of perfect success after an arbitrary number of iterations is calculted in two cases: the worst case and the average case. Finally the conclusions of this article is summerized in section 5.

\section{Review of Chang et al.'s Protocol}
\label{sec:1}

Throughout this section we briefly present a short review of Chang et al.'s Protocol. Alice and Bob have two secret binary strings 
$ ( a_1, a_2,\dots , a_n):= \mathrm{ID_{A}}$ and $ (b_1, b_2,\dots , b_u):= \mathrm{ID_{B}}$, already shared between themselves, which represent Alice's identity and Bob's one respectively.
Suppose that Alice wants to send Bob a secret binary message $(m_1, m_2,\dots , m_n):=M$. Then Alice and Bob persue the following procedure:

\begin{itemize}

\item[Step 1] Alice encrypts \textit{M} with $\mathrm{ID_{A}}$ using the simple XOR-operation and
obtains $(c_1, c_2, \dots , c_n):=C$, where $c_i= m_i+a_i \hspace{2mm}\mathrm{mod} \hspace{2mm}2$, for $i=1,\dots , n$.

\item[Step 2] According to \textit{C}, Alice creates \textit{n} qubits, called
$S_C$ in the manner: if a bit of \textit{C} is 0, she prepares the
corresponding qubit in $|0\rangle $ or $|+\rangle $ state at random; otherwise she prepares the corresponding 
qubit in $|1\rangle $ or $|-\rangle $ state randomly. According to $\mathrm{ID_B}$, Alice prepares $u$ qubits, called ${S_\mathrm{IDB}}$ as follows:
if a bit of $\mathrm{ID_B}$ is 0, she randomly prepares the qubit in $|0\rangle$ or
$|1\rangle$ state; otherwise she randomly prepares the qubit in $|+\rangle$ or
$|-\rangle$ state. Alice inserts ${S_\mathrm{IDB}}$ to ${S_{C}}$ randomly which forms a new binary sequence called 
${S_{C'}}$ and sends it to Bob.

\item[Step 3] After Bob receives ${S_{C'}}$, Alice publicly announces
the positions of ${S_\mathrm{IDB}}$ in ${S_{C'}}$. Then Bob extracts ${S_\mathrm{IDB}}$
and measures these photons in the correct bases according to
$\mathrm{ID_B}$. If a bit of $\mathrm{ID_B}$ is 0, he measures the corresponding qubit in $B_Z=\{|0\rangle, |1\rangle\}$; otherwise, $B_{X} =\{ |+\rangle, |-\rangle \} $ will be applied.

\item[Step 4] Bob announces the state of photons in ${S_\mathrm{IDB}}$ which he
received; the basis information is not included in this announcement.
For example, Bob uses bit 0 to denote state $|0\rangle$ and
$|+\rangle$, and 1 for $|1\rangle$ and $|-\rangle$. According to the above rule, Alice obtains the state of the
initial ${S_\mathrm{IDB}}$. Alice compares Bob's result with the state of
initial ${S_\mathrm{IDB}}$. If the error rate is low enough, Alice believes
that Bob is legal and no eavesdropping exists. In this condition,
the communication goes on; otherwise she interrupts
it. Alice and Bob discard the bits in ${S_\mathrm{IDB}}$, where the corresponding
photons in ${S_\mathrm{IDB}}$ are not received by Bob.

\item[Step 5] Alice publicly announces the bases of photons
in ${S_C}$. Bob measures ${S_C}$ in correct bases and obtains \textit{C}.

\item[Step 6] Bob decrypts \textit{C} with $\mathrm{ID_{A}}$ bit by bit using 
simple XOR-operation: $m_i=c_i + a_i \hspace{2mm} \mathrm{mod} \hspace{2mm} 2$, for $i=1,\dots , n.$ In other words, $M=C \oplus \mathrm{ID_B}$ (Note that, ``$\oplus $" is used to represent XOR-operation of two binary strings with same lengths).

\item[Step 7] Alice takes another $n$-bit binary string of secret
message, called $M_1$ and starts the next transmission.

\end{itemize}

\section{Description of The New Attack}
\label{sec:2}
As mentioned in previous section, the authentication inside the protocol is unidirectional, i.e. just Alice can verify Bob's identity and demonstrates his legitimacy. Therfore Bob cannot verify the sender's identity. Hence anyone can impersonate Alice and sends some arbitrary messages (indeed qubits) to Bob. First, we briefly explain the novel attack; the scenario of the attack is composed of two scenes:

\paragraph{First scene of the attack.}

Oscar prepares a binary sequence of length \textit{u}, such as $(e_1, e_2, \dots, e_u):={id_\mathrm{B}}$ (Note that $id_\mathrm{B}\neq \mathrm{ID_B}$ in general). In fact ${id_\mathrm{B}}$ is a candidate for Bob's identity binary string and changes after each session untill coincides (with high probability) on $\mathrm{ID_B}$. 

According to ${id_\mathrm{B}}$, Oscar creates a sequnece of qubits and obtains ${S_{id\mathrm{B}}}$ as follows: if a bit of ${id_\mathrm{B}}$ is 0, the corresponding qubit of ${S_{id\mathrm{B}}}$ is $|0\rangle$; otherwise, it is $|-\rangle$. Next he creates a random qubit sequence as $S_C$ and mixes it to ${S_{id\mathrm{B}}}$, obtaining ${S_{C'}}$. Then he sends ${S_{C'}}$ to Bob.

Invoking the protocol, after Bob receives ${S_{C'}}$, Oscar announces the positions of ${S_{id\mathrm{B}}}$ in ${S_{C'}}$. Then Bob measures the polarization of any photon of ${S_{id\mathrm{B}}}$ due to the corresponding bit of $\mathrm{ID_B}$. The rule is that he uses $B_Z$ basis, for corresponding ``0" bits and $B_X$ for ``1" bits.
Then Bob announces the state of photons in ${S_{id\mathrm{B}}}$ he
received. As mentioned at step 4 of the protocol, without lose generality, assume that Bob uses bit 0 to denote state $|0\rangle$ or
$|+\rangle$, and 1 for $|1\rangle$ or $|-\rangle$. 

In other words, if a bit of the string which announced by Bob is 0, it means that the corresponding qubit he received is either $|0\rangle$ or $|+\rangle$; otherwise, it is either $|1\rangle$ or $|-\rangle$. 

Thus, Oscar obtains the state of the
initial ${S_{id\mathrm{B}}}$. He compares Bob's result with the state of
initial ${S_{id\mathrm{B}}}$. If a bit of the string which announced by Bob, and the corresponding qubit of ${S_{id\mathrm{B}}}$ do not match, Oscar concludes that the corresponding bit of ${id_\mathrm{B}}$, say $e_i$ is wrong and changes it; otherwise the corresponding bit of ${id_\mathrm{B}}$ is probably correct, and the probabity of the correctness depends on the number of session iterations. By this manner, after each iteration a new $id_{\mathrm{B}}$ replaced by the previous one. If after \textit{k} iterations, no non-matching case is observed in a position, it means that the bit is correct with probability $1-2^{-k}$. Therfore, if remain \textit{t} matchings after \textit{k} iterations, the probability of coincident of ${id_\mathrm{B}}$ and $\mathrm{ID_B}$ will be $(1-2^{-k})^t$.

After Oscar obtains $\mathrm{ID_B}$ (with high enough probability), he can impersonate Bob. 

\paragraph{Second scene of the attack.}

Oscar intercepts the communication between Alice and Bob. Since Oscar has $\mathrm{ID_B}$, when Alice announces the positions of ${S_\mathrm{IDB}}$ in ${S_{C'}}$, Oscar measures the qubits in correct bases (with high enough probability). So Alice will be deceived, and the communication goes on. Then she announces the bases of photons in $S_C$. Therfore Oscar has \textit{C}, which is the message encrypted by $\mathrm{ID_A}$ using simple XOR-operation. Hence he can break it easily; see \cite{9} and refrences therein.

\section{Numerical Exmaples and Discussion}
\label{sec: 3}

It is clarified at step 4 of the protocol, that Alice considers the error rate when she compares Bob's result with the state of initial ${S_\mathrm{IDB}}$. If it is low enough, Alice verifies the legitimacy of the receiver; see section 2. Suppose that the error rate of the quantum channel is $\epsilon$, and ${id_\mathrm{B}}$ differs with $\mathrm{ID_B}$ in \textit{t} positions at the beginning of session described in the first scene of the attack. We show that after a few number of iterations, the first scene of the attack will be successful, considering $\epsilon < 0.02$ (and even in ideal cases with high probability). After \textit{k} iterations of the session, $\lceil (1-(3/4)^k).t\rceil$ wrong bits of ${id_\mathrm{B}}$, will be corrected, on average (since $(3/4)^{n} \longrightarrow 0$ and the number of iterations is a discrete quantity, the ceiling function is used). Also it will be clear to Oscar that each of the other $(u-t)$ bits is same as the correponding bits of $\mathrm{ID_B}$ with probability $(1-2^{-k})$.
So, after \textit{k} iterations, Oscar knows that every changed bit of latest $id_{\mathrm{B}}$ is correct, and the remaining \textit{x}-bit substring is probably the same as the corresponding substring of $\mathrm{ID_B}$. Hence after \textit{k} iterations, if \textit{x} bits do not change, ${id_\mathrm{B}}=\mathrm{ID_B}$ with probabilty $(1-2^{-k})^x$.
Table 1 shows the probability of equality ${id_\mathrm{B}}=\mathrm{ID_B}$ after \textit{k} iterations with several lengths of $\mathrm{ID_B}$.

\begin{center}

\begin{table}[h!]
\caption{the probability of coincident in the worst case with $k$ iterations.}
\label{tab: 1}
\begin{tabular}{cccc} 
\hline\noalign{\smallskip}
No. of iterations & &The length of $\mathrm{ID_B}$ \\
\textit{k} & (32-bit) & (64-bit) & (128-bit) \\ 
\hline\noalign{\smallskip}
10 & 96.9\% & 93.9\% & 88.2\% \\
11 & 98.4\% & 96.9\% & 93.9\% \\
12 & 99.2\% & 98.4\% & 96.9\% \\
13 & 99.6\% & 99.2\% & 98.4\% \\ 
\noalign{\smallskip}\hline
\end{tabular}

\end{table}
\end{center}

Note that table 1, shows the probability of succes in the worst-case for some number of iterations, i.e. the correction of wrong bits is not considered. But in general, the probabilty of success increases. Let \textit{u} be the length of $\mathrm{ID_B}$. Then there will be $t\leq u$ wrong bits in the first candidate $id_\mathrm{B}$. So, as mentioned above, after \textit{k} iterations of the session, $\lceil (1-(3/4)^k).t\rceil$ wrong bits, on average, will be corrected. We consider $t=u/2$. Table 2 shows the probability of success in the average-case.

\begin{center}

\begin{table}[h!]
\caption{the probability of coincident in the average case with $k$ iterations.}
\label{tab: 2}
\begin{tabular}{cccc} 
\hline\noalign{\smallskip}
No. of iterations & &The length of $\mathrm{ID_B}$ \\
\textit{k} & (32-bit) & (64-bit) & (128-bit) \\ 
\hline\noalign{\smallskip}
10 & 98.4\% & 96.8\% & 93.6\% \\
11 & 99.2\% & 98.4\% & 96.8\% \\
12 & 99.6\% & 99.1\% & 98.4\% \\
13 & 99.8\% & 99.6\% & 99.2\% \\ 
\hline\noalign{\smallskip}
\end{tabular}
\end{table}

\end{center}


\section{Modification of The Protocol}
\label{sec: 4}
As explained in the previous sections, Chang et.al.'s protocol may be broken by the proposed attack. In this section we expose a proposal to reinforce the protocol. Since the loophole of the protocol is originated from ``unidirectional" authentication, we propose a method for mutual authentication. This strategy can close the loophole and make the protocol more strong and resistant to the new attack. Since in the original protocol, it is assumed that Alice and Bob already sheared two secret strings $\mathrm{ID_A}$ and $\mathrm{ID_B}$, we will also suppose that Alice and Bob have common secrets containing $\mathrm{ID_A}:=(a_1,a_2, \dots , a_{2\ell} )$, $\mathrm{ID_B}:=(b_1,b_2, \dots , b_u )$ and secret key $\mathrm{K_{AB}}:=(k_1, k_2, \dots , k_n)$.  Also the authentication procedure and sending the secret message will be based on single photons and quantum memory as in the original protocol. The modofied protocol to send the message $M:=(m_1, m_2, \dots , m_n)$ is as follows:

\begin{itemize}
\item[Step 1] Alice encrypts the message $M$ by $\mathrm{K_{AB}}$ using XOR-operation, so that she obtains $C=M \oplus \mathrm{K_{AB}}$.
\item[Step 2] Exactly like the original protocol, Alice prepares $S_C$ and $S_{\mathrm{IDB}}$. In addition, She creates $\ell$ qubits, called $S_{\mathrm{IDA}}$, according to $\mathrm{ID_A}$. The procedure of creating $S_{\mathrm{IDA}}$ is that Alice randomly chooses $\ell$ bits of  $\mathrm{ID_A}$ as an ordered basis string by the rule: $0\equiv B_Z$ and $1\equiv B_X$. The remaining $\ell$ bits are used as an ordered qubit string measured (after reordering if neccessary) in the corresponding bases by the rule: $0 \equiv |0 \rangle $ or $ |+\rangle$ and $1 \equiv |1 \rangle $ or $ |-\rangle$ due to the corresponding basis. Then Alice inserts $S_{\mathrm{IDA}}$ and $S_{\mathrm{IDB}}$ randomly to  $S_C$, obtaining  $S_C'$, and sends it to Bob. 

\item[Step 3] After Bob receives  $S_C'$, Alice announces the positions of $S_{\mathrm{IDA}}$, \textit{t} bits of  $\mathrm{ID_A}$ used for sent qubits by order, and $S_{\mathrm{IDB}}$. Then Bob extracts the qubits of  $S_{\mathrm{IDA}}$ and  $S_{\mathrm{IDB}}$.

\item[Step 4] Bob measures the qubits of $S_{\mathrm{IDA}}$ and checks the legitimacy of Alice. If she is legal, then Bob measures the qubits of $S_{\mathrm{IDB}}$ and sends the result to Alice exactly like the original protocol; otherwise he interrupts it.

\item[Step 5] Alice cheks the leitimacy of Bob as described in \cite{8}. If he is legal, she announces the bases of $S_C$ qubits; otherwise she interrupts the communication.

\item[Step 6] Bob decrypts message \textit{C} by $\mathrm{K_{AB}}$, using XOR-operation to obtain the plaintext message \textit{M}.

\end{itemize}
\section{Security Analysis of The Modified Protocol}
\label{sec: 5}

As analyzed in \cite{8}, the original protocol is secure against most attacks such as man in the middle attack and quantum teleportation attack, but the proposed attack can break the protocol and the attacker can obtain $\mathrm{ID_{A}}$ and $\mathrm{ID_{B}}$ which are secrets already shared between Alice and Bob. The modification of the protocol which is proposed in the previous section, can fill the loophole of teh protocol. Since the loophole is emanated from unidirectional authentication, applying the bidirectional authentication proposed in the previous section fills it up. Formal speaking, suppose that attacker Oscar tries to impersonate Alice, and sends Bob a message. When Bob receives the message, Oscar should announces the positions of $\ell$ bits of $\mathrm{ID_{A}}$, which show his legitimacy. So he cannot impersonate Alice, unless he accesses to  $\mathrm{ID_{A}}$. Therfore any non-legal person who does not access to  $\mathrm{ID_{A}}$, cannot deceive Bob. On the other hand, if an adversary tries to impersonate Bob, after he receives the message, obtains no information about  $\mathrm{ID_{A}}$. More precisely, it is possible that an adversary receives the message, since Bob proves his legitimacy after Alice. But there is nothing to worry about, because when Alice announces the positions of the $\ell$ verification bits, she does not declare the correct bases of them. So the adversary obtains no information about  $\mathrm{ID_{A}}$. Moreover he cannot impersonate Bob (unless he accesses to  $\mathrm{ID_{B}}$). Therfore when Alice checks the legitimacy of the receiver, would not been deceived and interrupts the communication. In this situation, reordering of $\ell$ verification bits is neccessary to retry sending the message. To reach more scurity it is suggested that a derangment on $\ell$ verification bits be applied. Note that repetition of sending message to a non-legal person, allows him to guess $\mathrm{ID_{A}}$ applying a same method described in the first scene of the attack; see section 3. Hence, one can deduce if a qubit of $\mathrm{ID_{A}}$ loses once because of noisy channel or eavesdropping, it can be sent in the next transmissions safely, provided that its corresponding basis index, changes. However if a fixed qubit loses twice, it must be discarded. 

Clearly the modified protocol defeat man-in-the-middle attack and quantum teleportation attack as the original protocol can.

\section{Conclusion}
\label{sec: 6}
We have demostrated that Chang et al.'s protocol, is vulnerable to a specific attack, which is described in this article. Our attack scenario is divided into two scenes. Since Bob cannot check Alice's legitimacy, at the first scene, Oscar impersonates Alice. Then he chooses a candidate binary string for $\mathrm{ID_{B}}$, and corrects the wrong bits of it by sending a message several times to Bob, due to the original protocol. After afew iterations, he obtains $\mathrm{ID_{B}}$, with high enough probability. At the second scene, Oscar intercepts the transmission between Alice and Bob, and impersonates Bob easily. Then he can receive any message which is sent by Alice, measure the qubits in the correct bases, and finally obtain $\mathrm{ID_{A}}$ by breaking a simple XOR-encryption. Since the loophole of the protocol is originated from the unidirectional authentication, the mutual authentication is suggested to defeat the attack. Furthermore the modified protocol is secure against all attacks which the original one is.



\begin{thebibliography}{}
%
%




\bibitem{1} Bennett, C. H., Brassard, G. Quantum cryptography: Public key distribution and coin
tossing, in Proc. IEEE Conf. Computers Systems and Signal Processings (Bangalore,
India) December 10–19 (IEEE, New York, 1984), p. 175.


\bibitem{2} Hillery, M., Buzek, V. and Berthiaume, A., 1999. Quantum secret sharing.Physical
Review A, 59(3), p.1829.


\bibitem{3} Barnum, H., Crpeau, C., Gottesman, D., Smith, A. and Tapp, A., 2002. Authentica-
tion of quantum messages. In Foundations of Computer Science, 2002. Proceedings.
The 43rd Annual IEEE Symposium on (pp. 449-458). IEEE.




\bibitem{4} Long, G.L. and Liu, X.S., 2002. Theoretically efficient high-capacity quantum-key-
distribution scheme. Physical Review A, 65(3), p.032302.



\bibitem{5} Deng, F.G., Long, G.L. and Liu, X.S., 2003. Two-step quantum direct communica-
tion protocol using the Einstein-Podolsky-Rosen pair block.Physical Review A, 68(4),
p.042317.

\bibitem{r-6}Chang, Y., Xu, C., Zhang, S. and Yan, L., 2014. Controlled quantum secure direct communication and authentication protocol based on five-particle cluster state and quantum one-time pad. Chinese science bulletin, 59(21), pp.2541-2546.

\bibitem{r-5}Hassanpour, S. and Houshmand, M., 2015. Efficient controlled quantum secure direct communication based on GHZ-like states. Quantum Information Processing, 14(2), pp.739-753.

\bibitem{6} Deng, F.G. and Long, G.L., 2004. Secure direct communication with a quantum
one-time pad. Physical Review A, 69(5), p.052319.



\bibitem{7} Lucamarini, M. and Mancini, S., 2005. Secure deterministic communication without
entanglement. Physical review letters, 94(14), p.140501.


\bibitem{8} Chang, Y., Xu, C., Zhang, S. and Yan, L., 2013. Quantum secure direct commu-
nication and authentication protocol with single photons. Chinese Science Bulletin,
58(36), pp.4571-4576.



\bibitem{r-7} Hwang, T., Luo, Y.P., Yang, C.W. and Lin, T.H., 2014. Quantum authencryption: one-step authenticated quantum secure direct communications for off-line communicants. Quantum information processing,13(4), pp.925-933.


\bibitem{r-12} Yang, Y.Y., 2014. A quantum secure direct communication protocol without quantum memories. International Journal of Theoretical Physics, 53(7), pp.2216-2221.



\bibitem{r-9} Zou, X. and Qiu, D., 2014. Three-step semiquantum secure direct communication protocol. Science China Physics, Mechanics \& Astronomy,57(9), pp.1696-1702.


\bibitem{chou} Chou, Y.H., Zeng, G.J., Lin, F.J., Chen, C.Y. and Chao, H.C., 2014. Quantum Secure Communication Network Protocol with Entangled Photons for Mobile Communications. Mobile Networks and Applications, 19(1), pp.121-130.


\bibitem{r-8} Lai, H., Xiao, J., Orgun, M.A., Xue, L. and Pieprzyk, J., 2014. Quantum direct secret sharing with efficient eavesdropping-check and authentication based on distributed fountain codes. Quantum information processing, 13(4), pp.895-907.




\bibitem{r-2} Gao, G., 2014. Cryptanalysis and improvement of quantum secure communication network protocol with entangled photons for mobile communications. Physica Scripta, 89(12), p.125102.


\bibitem	{r-1} Ye, T.Y., 2015. Fault-tolerant authenticated quantum dialogue using logical Bell states. Quantum Information Processing, 14(9), pp.3499-3514.


\bibitem{r-3} Ye, T.Y., 2015. Quantum secure direct dialogue over collective noise channels based on logical Bell states. Quantum Information Processing,14(4), pp.1487-1499.


\bibitem{r-4} Xiao M, and Xu H W, 2015. High-Capacity Quantum Secure Communication with Authentication Using Einstein-Podolsky-Rosen Pairs. Chinese Physics Letters, (5), pp.1-4.





\bibitem{r-11} Hong-Yang, M., Guo-Qing, Q., Xing-Kui, F. and Peng-Cheng, C., 2015. Quantum network direct communication protocol over noisy channel. ACTA PHYSICA SINICA, 64(16).


\bibitem{r-13} Yan, C., Shi-Bin, Z., Li-Li, Y. and Gui-Hua, H., 2015. Controlled Deterministic Secure Quantum Communication Protocol Based on Three-Particle GHZ States in X-Basis. Communications in Theoretical Physics, 63(3), p.285.

\bibitem{r-14} Hu, J., Yu, B., Jing, M., Xiao, L., Jia, S., Qin, G. and Long, G., 2015. Experimental quantum secure direct communication with single photons. arXiv preprint arXiv:1503.00451.

\bibitem{r-10} Mi, S., Wang, T.J., Jin, G.S. and Wang, C., 2015. High-capacity quantum secure direct communication with orbital angular momentum of photons. Photonics Journal, IEEE, 7(5), pp.1-8.

\bibitem{r-15}  Uhlmann, A., 2016. Anti-(conjugate) linearity. Sci China-Phys Mech Astron, 59(3): 630301.




\bibitem{r-16} Li, T. and Yin, Z.Q., 2016. Quantum superposition, entanglement, and state teleportation of a microorganism on an electromechanical oscillator. Science Bulletin, pp.1-9.












\bibitem{9} Schneier, B., 2007. Applied cryptography: protocols, algorithms, and source code in
C. John Wiley \& sons.

\end{thebibliography}


\end{document}